%% file: article.tex
\documentclass[12pt]{article}
\usepackage{epsfig}

\input econfmacros.tex
\def\Title#1{\begin{center} {\Large {\bf #1} } \end{center}}

\begin{document}

\Title{Polarization in GRB Afterglows}

\bigskip\bigskip


\begin{raggedright}  

{\it Gunnlaugur Bj\"ornsson\index{Bj\"ornsson, G.}\\
Science Institute\\
University of Iceland\\
Dunhagi 3, 107 Reykjavik, Iceland}
\bigskip
\end{raggedright}

\section{Introduction}

Over the past five years a clear and consistent picture of the cause of 
Gamma-Ray Burst (GRB) afterglows has emerged. The afterglow is thought to 
originate in a decelerating relativistic fireball and is generally believed 
to be synchrotron emission (see e.g. \cite{piran00,mesz02} for a review). 
Synchrotron emission can, under favorable conditions, be up to 
$60-70\%$ polarized (e.g.\ \cite{ryblight}). It is therefore natural to 
expect that the GRB afterglow emission might show some degree of polarization. 
The amount of polarization may depend on the magnetic field structure and its
regularity, the geometry of the collimated outflow and on the properties of the 
local burst environment. It is important for our understanding of the GRB 
phenomenon, to be able to account properly for all polarization from 
these sources. It may, in particular, provide important information on the 
geometry and magnetic field properties of the bursts.   

\section{Models}

Several models have been suggested in order to quantify the polarization
that might potentially be observed from bursts. In addition to the intrinsic 
afterglow polarization, it is also very important to account for the 
polarization that might originate in our Galaxy as well as in the host galaxy
of a given burst. We will briefly discuss these possibilities in turn.\\
{\bf i) Coherent magnetic patches:} The relativistic shock is composed of 
several causally disconnected patches or domains, each patch having a regular 
magnetic field structure and exhibiting a polarization level up to 
$P_{\rm max}\sim 60-70\%$. An observable collection of $N$ patches would 
result in a net polarization of $P\sim P_{\rm max}/\sqrt{N}$, with the size of 
the patches growing at the speed of light \cite{gruzwax99}.\\
{\bf ii) Microlensing:} If a burst is microlensed a particular magnetic patch
might get briefly magnified resulting in a relatively short episode of variable
polarized emission \cite{loebpern98}. The degree of polarization at maximum 
depends on the number of magnified patches.\\
{\bf iii) Interstellar scintillation:} is also able to produce polarization 
in the radio of the order of $10-20\%$, about a week after the 
burst \cite{medloeb99}.\\
{\bf iv) Collimated fireballs:} Recent evidence suggests that the relativistic
outflow associated with a GRB is collimated. If the observer's line of sight
happens to coincide with the outflow symmetry axis, no net polarization is expected. 
If the line of sight, on the other hand, makes an angle with the collimation axis, 
the symmetry is broken, and a net polarization may arise. It can be shown 
\cite{ghislaz99, sari99} that there will be at least two peaks in the 
polarization light curve, the first one arising when the observer only receives
radiation from that part of the relativistic beam that is inside the outflow 
geometry. The missing part therefore does not contribute to the total emission
and a net polarization arises \cite{ghislaz99}. The second peak occurs around 
the break time in the optical light curve and reflects the maximum asymmetry 
between the emitting areas contribution to the different polarization directions 
(\cite{ghislaz99, sari99}). A third peak may arise if the jet is spreading 
laterally~\cite{sari99}, but this occurs typically a week after the burst, 
at which time the flux level may already be too low to allow a reliable 
polarization measurement. An interesting and testable prediction of the models 
is that the polarization will decrease to zero in between the first two peaks, 
accompanied by a $90^\circ$ change in the polarization angle as the degree of 
polarization starts increasing again. The collimated outflow model has also 
been considered in the case of an observer with a line of sight outside the 
initial opening angle of the jet \cite{granot02}. A variant of the model, 
with a bright emission core near the outflow axis surrounded by dimmer wings, 
results in only one polarization maximum and no change of polarization 
angle \cite{rossi03}. It is interesting to note that this variant of the model
predicts maximum polarization at the light curve break time, whereas the 
original models with a homogeneous outflow predicts a minimum polarization level
at that time.\\
{\bf v) Other:} In the case of weakly polarized sources it is important to
properly account for the Galactic interstellar polarization. In addition, 
some degree of polarization may similarly be induced in the host galaxy.
In the Galactic case, correcting for this ideally requires observing stars 
of low intrinsic polarization that are sufficiently distant to ensure an 
essentially full probe of the interstellar medium, but also sufficiently near 
the burst location on the sky to allow a meaningful subtraction of the 
polarization due to the interstellar medium. Alternatively, the interstellar 
polarization may be estimated from the color excess, $E(B-V)$,
that contributes a maximum of $P(\%)=9E(B-V)$ \cite{serko75}, depending on 
the line of sight through the Galaxy. Hence, a color excess of at least 
$E(B-V)\sim 0.3$ is needed to produce interstellar polarization levels of 
the order of $3\%$. For most cases to date extinction maps towards bursts 
imply an interstellar contribution to the polarization up to about 1\% at 
most. Similar correction may be necessary in the host galaxy of the burst, 
but this is at present much more difficult to estimate. Employing Galactic 
analogies, one may try to use similar approach as in our own Galaxy and 
estimate the host polarization contribution from the host extinction, if 
this can be inferred. 
\begin{table}[bht]
\begin{center}
\begin{tabular}{l|rrrrr}  
GRB     & P(\%)          & PA (deg.)        & T  & Ref.   \\ \hline
980425  & 0.6            & $80$             & $\sim  8$    & \cite{patat01}  \\
980425  & 0.4            & $67$             & $\sim 25$    & \cite{patat01}  \\
980425  & 0.53           & $49$             & $\sim 42$    & \cite{kay98}    \\ \hline
990123  & $<2.3$         &                  & 0.76       & \cite{hjorth99} \\ \hline
990510  & $1.7\pm 0.2$   & $101\pm 3$       & 0.77       & \cite{covino99} \\
990510  & $1.6\pm 0.2$   & $98\pm  5$       & 0.86       & \cite{wijers99} \\ \hline
990712  & $2.9\pm 0.4$   & $121.1\pm 3.5$   & 0.44       & \cite{rol00}    \\
990712  & $1.2\pm 0.4$   & $116.2\pm 10.1$  & 0.70       & \cite{rol00}    \\
990712  & $2.2\pm 0.7$   & $139.2\pm 10.4$  & 1.45       & \cite{rol00}    \\ \hline
010222  & $1.36\pm 0.64$ &                  & 0.94       & \cite{bjor02}   \\ \hline
011211  & $<2.7$         &                  & 1.5        & \cite{covino02a} \\ \hline
020405  & $1.5\pm 0.4$   & $172\pm 8$       & 1.2        & \cite{masetti02} \\
020405  & $9.9\pm 1.3$   & $-0.1\pm3.8$     & 1.3        & \cite{bersier03} \\
020405  & $1.96\pm 0.33$ & $154\pm 5$       & 2.18       & \cite{covino02b} \\
020405  & $1.47\pm 0.43$ & $168\pm 9$       & 3.27       & \cite{covino02b} \\ \hline
020813  & $1.8-2.4     $ & $153-162 $       & 0.2-0.3    & \cite{barth03}   \\
020813  & $0.80\pm 0.16$ & $144\pm 6$       & 0.58       & \cite{covino02c} \\ \hline
\end{tabular}
\caption{Polarization in optical afterglows. P is the degree of polarization, PA
is the polarization angle and T is the time in days after the gamma ray event.}
\label{table}
\end{center}
\end{table}
%
Available host extinction estimates (e.g.\ \cite{holland02, klose00}) indicate 
that a range of polarization levels of this origin may be expected.
The contribution to the polarization in the host and the 
interstellar Galactic medium is however expected not to vary with time. 
Therefore, observing variable polarization may be taken as an indication of 
an origin intrinsic to the afterglow.

\section{Observations}

Measuring the polarization in a GRB afterglow was first attempted in the
case of GRB~990123~\cite{hjorth99} and resulted in an upper limit of $2.3\%$.
The first positive detection was for GRB~990510~\cite{covino99, wijers99},
and since then polarized emission has been measured in about 5 other afterglows.
The observations are summarized in Table~\ref{table}. Positive detections have 
been published in 5 cases and upper limits in 2 cases. Polarization was also 
measured for SN1998bw/GRB~980425, but these data were obtained at least a week 
or more from the burst at levels that are a factor of few lower than in other 
bursts. The polarization level in this case is similar to that observed in 
other supernovae and is more likely to originate in the non-symmetric expansion 
of the supernova ejecta. 

In addition to the bursts listed in Table~\ref{table}, there are indications 
of polarized emission from GRB~021004, that occurred after the NBSI, but the 
details of most of these measurements are still unpublished 
(e.g.\ \cite{covino02d, rol02, wang03}).

\section{Discussion}

Existing afterglow models predict a modest to strong polarization level of 
$P\approx 10-20$\%. In addition, some models based on collimated outflows 
predict a $90^\circ$ change in polarization angle between two adjacent maxima 
in the polarization light curve. 

Most measurements to date, however, indicate a low polarization level. In 
all cases but one is the polarization $<3$\% at all times. In addition, the 
variation in polarization angle within an individual burst is small and never 
close to the 90$^\circ$ change predicted by the collimated outflow models. 
In most cases is the angle in fact consistent with being constant. 
If conditions in the host galaxies are similar to the conditions in our
Galaxy, we may expect low polarization levels within the host. It has also 
been argued that scattering within the host would not be able to produce a 
few percent polarization within a day of the burst as commonly observed. 
The path length difference 
between the scattered (polarized) light and the direct light required to
produce a few percent polarization would cause a delay in the polarized light 
by weeks to months \cite{wijers99}. 

One measurement of GRB~020405 shows an exceptionally high polarization of almost 
$10$\%. This value is obtained a couple of hours following a measurement
at the 1.5\% level. The strong variation indicates an origin intrinsic to the 
afterglow rather in the interstellar medium, but it also makes it very difficult 
to account for using the collimated outflow models \cite{ghislaz99, sari99}, 
in particular as no 
break was observed in the light curve for the first few days. Alternative 
explanations may be invoked \cite{bersier03}, that do not require a break 
in the light curve \cite{gruzwax99, loebpern98}.

We conclude that the polarization at the 
few percent level observed in a number of optical afterglows is most likely 
intrinsic to the source. Our theoretical understanding of it is, however, 
still fragmentary and far from complete. \\

Polarization is becoming routinely measured in optical afterglows and it is 
of crucial importance to sample the polarization light curve as densely as 
possible within the first 2-3 days following a burst as this may provide very 
strong constraints on the afterglow models. It may also provide detailed 
information on the properties of the local burst environment. Spectropolarimetry 
is also proving to be an invaluable tool that is already indicating that the 
polarization may be both wavelength and time dependent \cite{barth03, wang03}.
We do, however, need better theoretical modeling to advance our understanding 
of the polarization properties of the afterglows.

\bigskip
This work was supported in part by the University of Iceland Research Fund and
a {\em Special grant} from the Icelandic Research Council.

\def\Discussion{
\setlength{\parskip}{0.3cm}\setlength{\parindent}{0.0cm}
     \bigskip      {\Large {\bf Discussion}} \bigskip}
\def\speaker#1{{\bf #1:}\ }
\def\endDiscussion{}

\Discussion

\speaker{J.\ Trier Frederiksen} Are there any models involving 
polarization by interaction with interstellar medium?

\speaker{G.\ Bj\"ornsson} The basic model is that of scattering 
of dust grains. The models I have discussed that are related 
to $E(B-V)$, account for interstellar polarization in our Galaxy and
can in principle be applied to the host galaxy contribution too.
In most cases, however, is the host extinction rather small,
too small in fact to fully explain the observed polarization level.
Polarization variability is also an indication of a different origin.

\speaker{S.\ Kulkarni}
I believe that polarization of the order of $1-3\%$ towards stars
is not uncommon.  Likewise one may expect some degree of
polarization induced by the interstellar medium within the
host galaxy. Thus the modeling should include a (unknown)
foreground polarization (i.e.\ due to reddening of the afterglow).
This may severely limit the conclusions drawn for the few
(1-3) data points. Indeed, as you noted the polarization for
several afterglows was found to be constant and a few per cent
(except 020405).

\speaker{G.\ Bj\"ornsson} Yes, I agree. The only way I see to 
distinguish between these polarization sources is to sample the 
polarization light curve sufficiently densely in addition to
spectropolarimetry. This would allow us to monitor variability in 
the polarization levels as well as changes in the polarization 
angle that has until now appeared basically constant. It would in
particular allow us to follow rapid changes in polarization as was 
observed in 020405. This however, requires a dedicated effort and
given the faintness and rapid fading of the afterglows, will most 
likely require a combination of a bright burst and favorable 
observational circumstances.

\endDiscussion
 
\end{document}

%% file: econfmacros.tex
\def\beq{\begin{equation}}
\def\eeq#1{\label{#1}\end{equation}}
\def\eeqn{\end{equation}}

\def\beqa{\begin{eqnarray}}
\def\eeqa#1{\label{#1}\end{eqnarray}}
\def\eeqan{\end{eqnarray}}

\let\bar=\overbar

\def\Dslash{\not{\hbox{\kern-4pt $D$}}}
\def\dslash{\not{\hbox{\kern-2pt $\del$}}}

\def\msb{{\bar{\ssstyle M \kern -1pt S}}}

